\documentclass[10pt, journal, twocolumn]{IEEEtran} 
\usepackage{amssymb,amsmath}
\usepackage{cite}
\usepackage{tabularx}
\usepackage{graphicx,subfigure}
\usepackage{psfrag}
\usepackage{url}
\usepackage{hyperref}
\usepackage{booktabs}
\usepackage{lscape}
\usepackage{multirow}
\usepackage{tablefootnote}
\usepackage[absolute,overlay]{textpos}
\usepackage{color}
\usepackage{colortbl}
\definecolor{lightgray}{gray}{0.95}
\definecolor{lightgray2}{gray}{0.8}
\usepackage{makecell}
\usepackage{multirow}
\usepackage{array}
\newcolumntype{L}[1]{>{\raggedright\let\newline\\\arraybackslash\hspace{0pt}}m{#1}}
\newcolumntype{R}[1]{>{\raggedleft\let\newline\\\arraybackslash\hspace{0pt}}m{#1}}
\usepackage{arydshln}

\usepackage{verbatim}

\begin{document}
	
    \title{Toward Proactive RF Charging Scheduling: Generative AI for Decision Support}
    \author{
        \IEEEauthorblockN{
        Amirhossein  Azarbahram,~\IEEEmembership{Graduate~Student~Member,~IEEE},
        Osmel M. Rosabal,~\IEEEmembership{Member,~IEEE},\\
        David Ernesto Ruiz-Guirola,~\IEEEmembership{Member,~IEEE},
        Melike~Erol-Kantarci,~\IEEEmembership{Fellow,~IEEE},
        Kaibin~Huang,~\IEEEmembership{Fellow,~IEEE},\\ 
        Onel L. A. L\'opez,~\IEEEmembership{Senior Member,~IEEE}
        }
    \thanks{Amirhossein  Azarbahram, Osmel M. Rosabal, David Ernesto Ruiz-Guirola, and Onel L. A. L\'opez are with the Centre for Wireless Communications (CWC), University of Oulu, Finland (e-mails: \{amirhossein.azarbahram, osmel.martinezrosabal, david.ruizguirola, onel.alcarazlopez\}@oulu.fi). 
    Melike~Erol-Kantarci is with the School of Electrical Engineering and Computer Science, University of Ottawa, Ottawa, Canada (e-mail: {melike.erolkantarci}@uottawa.ca). Kaibin Huang is with the Department of Electrical and Electronic Engineering, The University of Hong Kong, Hong Kong (e-mail: huangkb@eee.hku.hk).}
    \thanks{This work is partially supported by the Research Council of Finland (Grants 348515, 369116 (6G Flagship), and 362782 (ECO-LITE)); by the European Commission through the Horizon Europe/JU SNS project AMBIENT-6G (Grant 101192113); by the Research Grants Council of the Hong Kong Special Administrative Region, China under a fellowship award (HKU RFS2122-7S04), NSFC/RGC CRS (CRS-HKU702/24); and by the Canada Research Chairs program.}}
    \maketitle

\begin{abstract}
Radio frequency wireless power transfer (RF-WPT) is an enabling technology for supporting uninterrupted communications in future Internet of Things systems by reducing the need for battery replacement and mitigating battery-waste-related issues. For large-scale RF-WPT deployment, one of the main challenges is the scheduler-level resource allocation. Specifically, the RF charger must decide how much energy to deliver, when, and to whom, under limited charging resources, incomplete receiver-side information, and uncertain near-future charging conditions. This article positions generative artificial intelligence (GenAI) as a promising tool for this setting because it can foresee multiple plausible charging scenarios conditioned on coarse operational context and receiver-side information. We propose GenAI to act as an uncertainty-aware support layer for the RF-WPT scheduler rather than as a standalone forecasting or decision-making tool. To this end, we first revisit the main challenges of RF-WPT scheduling, and discuss how major GenAI families can support uncertainty-aware charging decisions by generating scenario-based inputs for downstream tasks. We then present a case study showing that distribution-aware prediction can improve robust charging decisions over deterministic, ensemble, and non-learning baselines, particularly under risk-sensitive objectives. Finally, we outline key open challenges and future research directions.
\end{abstract}


\section{Introduction}\label{intro}

The rapid growth of the Internet of Things (IoT) is driving the deployment of massive numbers of low-power devices, many of which operate under stringent energy constraints in time-sensitive use cases. For this, battery-less operation has emerged as an attractive alternative to conventional battery-powered deployments, whose large-scale operation is difficult to sustain due to the environmental impact and the cost of manually replacing or recharging batteries. This has intensified interest in energy harvesting (EH) technologies from both ambient and dedicated sources \cite{ZedHEXA}.

Among dedicated EH technologies, radio frequency wireless power transfer (RF-WPT) stands out due to its inherent multi-user charging support and potential operation in non-line-of-sight conditions. More importantly, RF-WPT can be partially enabled by existing wireless communications infrastructure, facilitating its integration as an enabling service of future IoT networks. Its practical relevance is also increasing as standardization and early commercial efforts begin to emerge, particularly for low-power IoT devices, battery-free sensing, and smart infrastructure applications.\footnote{See, e.g., the AirFuel Alliance RF wireless charging standard, Powercast wireless power solutions for retail and IoT, and Wiliot ambient IoT platforms.}
However, in dense IoT deployments, RF-WPT is not only about transferring power wirelessly but also about deciding where, when, to whom, and how much power to deliver under limited charging resources, thereby giving rise to the RF-WPT scheduling problem \cite{AIOTsurvey}.

Scheduling becomes even more critical as wireless systems are becoming increasingly context-aware, enabling more intelligent sensing, control, and resource management. As device operation adapts to context, energy demand becomes more heterogeneous across the network. Thus, addressing the scheduling problem requires moving beyond purely model-based designs, such as rule-based policies, periodic transmission, and deterministic optimization, toward data-driven approaches that can exploit contextual information about the network. This need arises because device behavior, environmental dynamics, and propagation conditions are often difficult to model explicitly, event-driven, and only partially observable. In RF-WPT systems, contextual information can be obtained from observed historical demand patterns and auxiliary descriptors exchanged during network operation, enabling charging decisions to better align with device operational behavior. However, the stringent energy limitations of the devices prevent frequent reporting of their current energy state and charging needs, while unexpected energy outages may occur before devices are scheduled for charging. Consequently, RF chargers may operate with incomplete or outdated information, further complicating charging scheduling.

Recent studies have explored generative artificial intelligence (GenAI) in related domains, suggesting that it may offer a useful set of tools for this type of decision problem. {Throughout this work, we use GenAI broadly to refer to models that learn structured data distributions and can generate new samples or outputs from them.} In EH IoT networks, GenAI has been discussed as a broad enabler for forecasting, optimization, and network management \cite{xie2025ehiot}. In wireless-network optimization, GenAI has been positioned as a tool for offline exploration and diverse scenario generation for proactive resource allocation \cite{khoramnejad2025xg}. In RF sensing, GenAI can be used to recover, complete, or synthesize radio-environment information under incomplete observations \cite{wang2025generative}. More broadly, GenAI has also been framed as a general foundation for future communication systems \cite{nguyen2026generative}. These works collectively highlight the value of GenAI for uncertainty modeling and decision support, but they fall short of formulating RF-WPT charging as a scheduler-level problem in which generated scenarios must be directly translated into charging decisions.

This gap is important because conventional predictive models typically reduce uncertainty to a single forecast, e.g., by predicting one receiver-wise future demand or average charging-demand trajectory over the scheduling horizon. This is a serious limitation in RF-WPT scheduling, since the same observed context may still lead to several plausible future charging patterns. In such cases, the scheduler benefits not only from knowing the most likely future but also from understanding the structure and likelihood of other plausible outcomes. This is precisely where generative models become relevant, as they can represent and sample from a conditional distribution over future evolutions. {Feedback-driven approaches such as reinforcement learning are also attractive
in principle, but they may be difficult to deploy when frequent interaction
or real-time feedback is unavailable. GenAI likewise requires sufficiently
representative training data, although it can be trained offline without
continual interaction, while online adaptation may still be beneficial.}

\begin{figure*}
    \centering
    \includegraphics[width=0.95\linewidth]{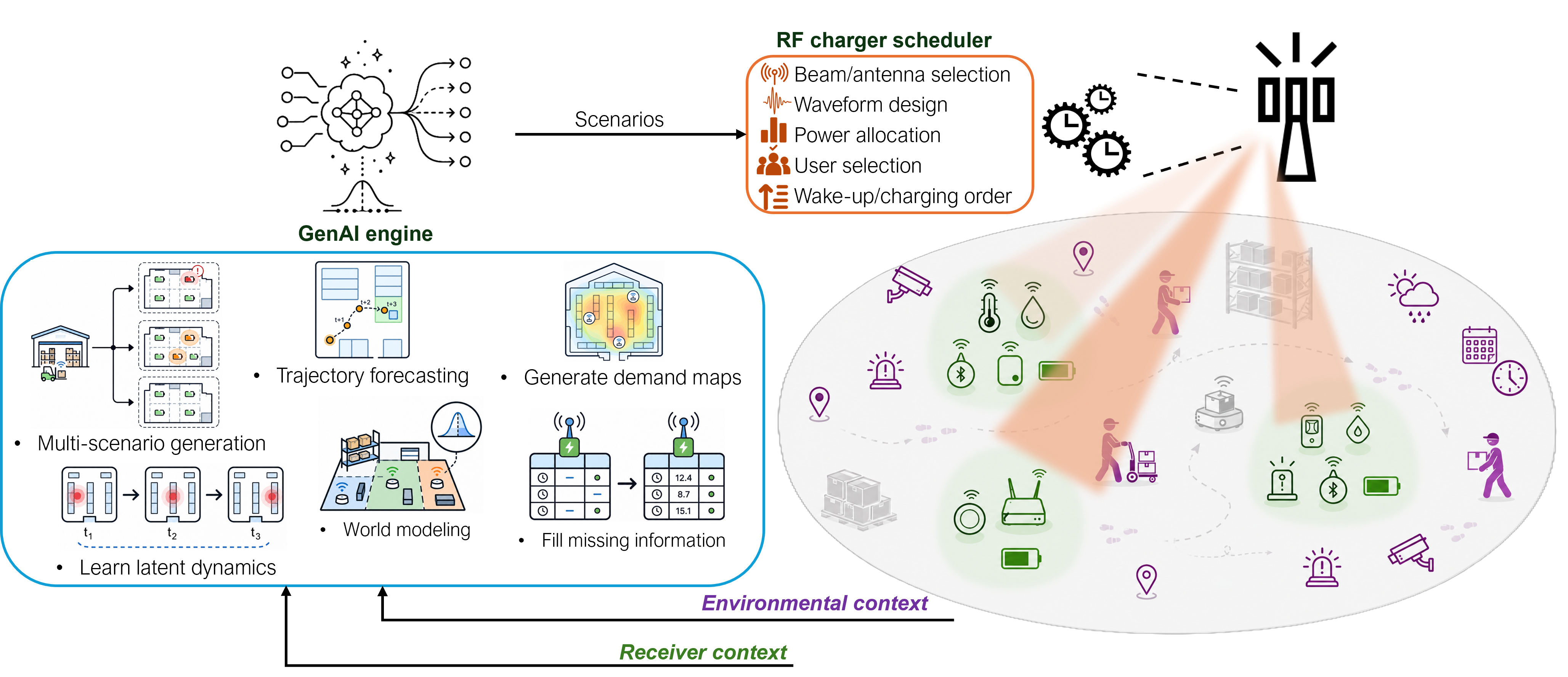}
    \caption{Vision toward intelligent GenAI-assisted RF-WPT charging scheduling.}
    \label{fig:vision}
\end{figure*}

Here, rather than treating GenAI as a standalone forecasting/predictive tool, we position it as an uncertainty-aware support layer for the RF-WPT scheduler in multi-receiver IoT networks. To the best of our knowledge, this is the first work to explicitly investigate GenAI-assisted, scenario-aware RF-WPT scheduling, in which the central goal is to generate plausible future charging conditions and expose them in a form that the RF charger can directly exploit when distributing limited charging resources across receivers. This perspective is particularly relevant to battery-constrained and battery-less IoT devices, whose operation depends strongly on the timely availability of energy. In such settings, instantaneous transfer efficiency and current battery state alone are insufficient for scheduling, because the RF charger must act under incomplete information and uncertain near-future conditions. Fig.~\ref{fig:vision} illustrates this overall vision, where a GenAI engine processes contextual information and receiver-side observations to support scenario-aware RF-WPT scheduling and downstream charging decisions. The main contributions are as follows:
\begin{itemize}
    \item We revisit RF-WPT from a scheduling-centric perspective and identify the main sources of uncertainty and context dependence limiting conventional charging policies.
    \item We establish a GenAI-based perspective for uncertainty-aware charging scheduling by analyzing major GenAI approaches and identifying how their generative capabilities can support scheduling in RF-WPT systems.
    \item We present a practically motivated case study showing how generative demand scenarios can improve robust RF-WPT charging decisions compared with deterministic, ensemble, and simple non-learning baselines.
    \item We identify key open challenges and highlight promising directions for future research.
\end{itemize}

\section{RF-WPT scheduling}

The low end-to-end efficiency of RF-WPT systems has historically driven research toward maximizing energy transfer performance \cite{Rosabal.2023}. Existing approaches overcome distance-induced propagation losses by deploying multiple RF chargers and by designing beamforming techniques to efficiently focus energy toward the devices \cite{Rosabal.2023,AIOTsurvey}. In parallel, the characteristics of EH circuits have been exploited through waveform design. Consequently, existing system-level design strategies inherently rely on network-side information, including device locations, channels, and/or detailed EH circuit models \cite{Rosabal.2023}. 

While effective for meeting short-term charging demands, this prevailing perspective overlooks the increasing complexity of emerging IoT ecosystems. In fact, the shift toward goal-oriented operation and the proliferation of different device types often result in highly varying energy demands \cite{ZedHEXA}. Consequently, charging must be treated as a spatiotemporal resource allocation problem. In essence, future RF-WPT will require novel policies to match more complicated future scenarios beyond merely satisfying immediate energy needs.

In a multi-user setting where devices activate at/for very different times, scheduling becomes particularly challenging, as variations in channel conditions and EH circuit parameters lead to different charging rates. Moreover, energy usage protocols, e.g., harvest-use, harvest-store-use, or harvest-use-store, impose additional constraints that must be explicitly incorporated into scheduling decisions. Moreover, each activation cycle for a given device may trigger different functionalities, causing variations in the required energy on the same device across multiple cycles. As the number of devices grows, the scheduling problem becomes not only inherently combinatorial, due to the joint optimization of user selection, power, beamforming, and time allocations, but also difficult to scale. In fact, allowing devices to frequently request charging services is impractical. Such signaling consumes a non-negligible portion of their limited energy budget and can lead to excessive access congestion when many devices attempt to communicate with the network simultaneously. 

Efficient RF-WPT operation under the aforementioned challenges calls for strategies that infer the environmental structure ahead of time. In this regard, learning and exploiting spatiotemporal patterns in device activity, historical energy usage, application-driven requirements, and environmental dynamics becomes essential. By leveraging such contextual information, the network can proactively infer when, where, how, and to whom energy will be required, thereby enabling sustained device operation with minimal signaling overhead.

All in all, RF-WPT scheduling is characterized by uncertainty, high dimensionality, and strong context dependence. The RF charger must plan the energy delivery while accounting for incomplete or outdated channel information, nonlinear EH behavior, varying receiver types, and fluctuating demands.

\section{Generative AI for charging scheduling}




In many RF charging scenes, the same observed context can lead to multiple plausible future charging patterns, limiting the effectiveness of deterministic rule-based schedulers, particularly during rare but important events. This highlights the need for GenAI, which can learn from data distributions, generate scenarios, and expose the scheduler to various plausible outcomes instead of relying on a single forecast~\cite{xie2025ehiot,khoramnejad2025xg}.

\subsection{GenAI Models}
Here, we focus on four representative GenAI model families relevant to RF-WPT scheduling. Variational autoencoders (VAEs) and conditional VAEs are suitable for learning compact latent representations of spatiotemporal charging and demand patterns, enabling efficient scenario generation and uncertainty-aware prediction~\cite{VAE_main,xie2025ehiot,zhang2025ioev}. Generative adversarial networks (GANs) are useful for synthesizing realistic charging traces, channel samples, or rare event patterns when measured data are scarce or imbalanced~\cite{GAN_main,wang2025generative}. Diffusion models generate high-fidelity, diverse samples through progressive denoising, making them suitable for generating plausible future charging maps and preserving multimodal uncertainty~\cite{diffusion_main,xie2025ehiot,zhang2025ioev,khoramnejad2025xg}. Finally, Transformer-based LLMs are relevant when charging decisions depend on long temporal dependencies, heterogeneous contextual signals, or multi-modal inputs, supporting contextual understanding, anomaly detection, and decision-making~\cite{LLM_main,TRM_main,wang2025generative,zhang2025ioev}. 

{From a deployment perspective, these families differ notably in
computational and inference cost. CVAEs and GANs generally enable
relatively lightweight sampling, whereas diffusion models require
iterative denoising and may be better suited to less
latency-sensitive scheduling unless accelerated sampling is used.
LLMs typically impose higher memory and inference costs and are thus more
appropriate for higher-level orchestration than real-time charging
control. Exact inference and computational cost depend on the
specific architecture and model size, application dimensionality,
sampling configuration, hardware platform, and number of generated
scenarios.}

\subsection{Potential Roles in RF-WPT Charging Scheduling}
{Table~\ref{tab:genAI_table}} further illustrates the principles of GenAI model families and how they can contribute to charging scheduling. It also summarizes the role of the GenAI models from a charging scheduling perspective, outlining their contributions to scenario generation, uncertainty modeling, synthetic data creation, context interpretation, and decision support with scarce or incomplete observations.

\begin{table*}[t]
    \centering
    \caption{Potential roles of major GenAI families in RF-WPT charging scheduling.}
    \label{tab:genAI_table}
    \includegraphics[width=0.98\linewidth]{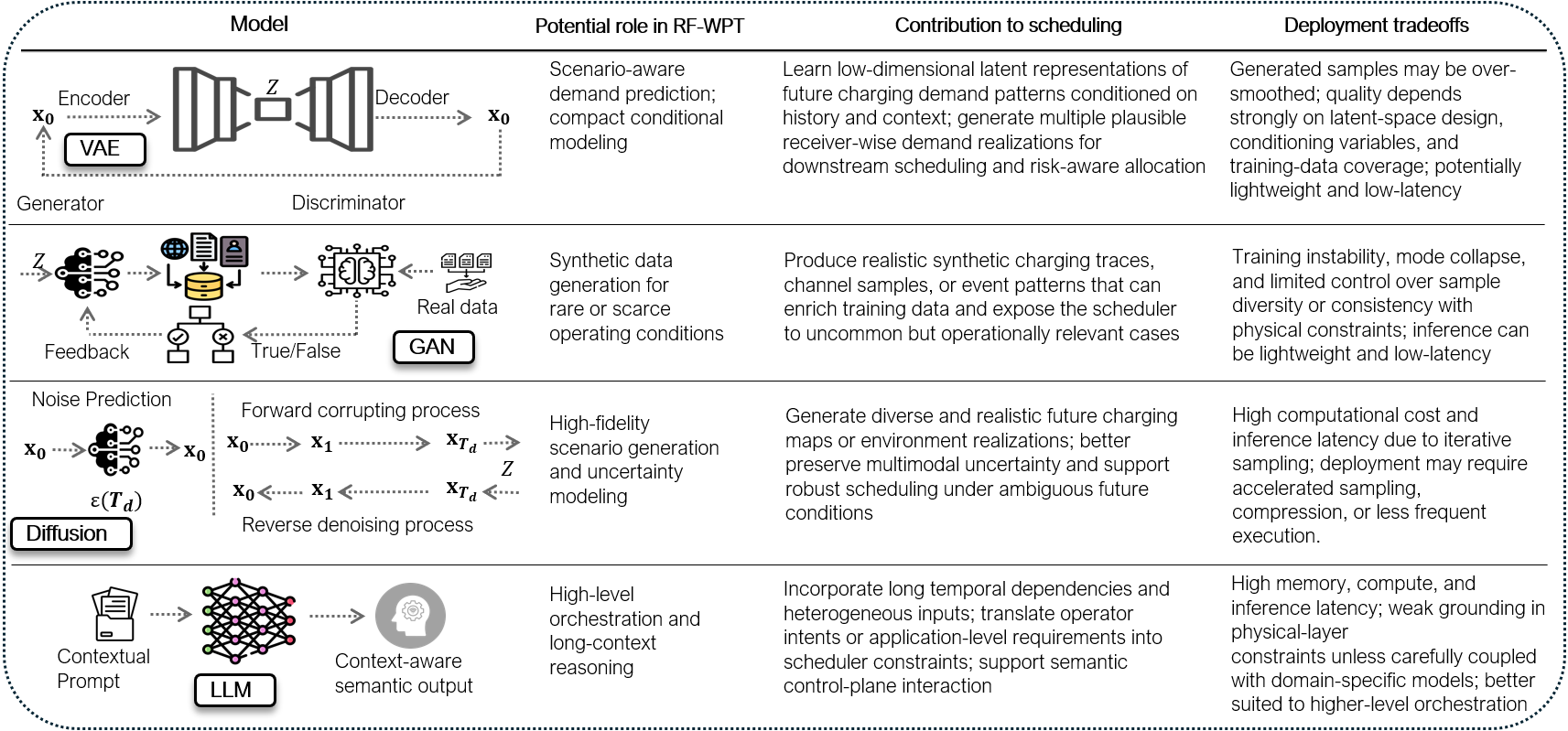}
\end{table*}

In this regard, a promising use of GenAI in RF-WPT is policy support through scenario-aware demand prediction. Rather than replacing the scheduler, GenAI can act as a learned model that provides candidate futures or synthetic experiences to a downstream optimizer or learning agent. Instead of outputting a single future charging vector, a generative predictor can sample multiple plausible receiver-wise demand realizations over a scheduling horizon~\cite{nguyen2026generative}. By exposing the scheduler to a distribution of futures, GenAI enables robust allocation against unfavorable yet plausible outcomes. In practice, this allows the decision module to explicitly optimize and plan for risk-sensitive objectives. In addition, synthetic but structured samples can support training when extensive online interaction or representative transition data are difficult to obtain. Hence, hybrid designs in which GenAI feeds a model-predictive controller/policy can support better scheduling both by generating better predictions and by preserving uncertainty in a form that can be directly exploited by the optimizer~\cite{xie2025ehiot,adam2025gai,khoramnejad2025xg}. 

Charging performance is influenced not only by demand but also by channel conditions, blockage, geometry, antenna orientation, and the nonlinear RF-to-direct current conversion characteristics. Since these variables are difficult to observe in real deployments, GenAI can be used to understand hidden patterns in propagation and the operating environment, reconstruct missing channel states, generate synthetic yet realistic radio environments, or infer latent spatial patterns that influence EH efficiency. This opens the door to GenAI-assisted beam selection, waveform-aware scheduling, and placement-aware charging decisions. For example, the scheduler may use generative channel samples to evaluate whether a narrow high-gain beam should be assigned to a critical receiver, or whether a broader multi-receiver strategy is preferable under channel uncertainty. Similarly, generated environment states can support proactive decisions in the presence of likely blockage, mobility, or task-induced demand shifts~\cite{khoramnejad2025xg,xie2025ehiot}.


At the control-plane level, GenAI may support higher-level orchestration by translating operator intents or application-level charging requirements into scheduler constraints. For instance, service directives such as prioritizing safety-critical nodes, preserving minimum energy neutrality, or meeting freshness targets can be converted into machine-readable objectives and then enforced by the lower-layer charging scheduler. While such language-driven control is not a replacement for physical-layer intelligence, it could simplify interaction with complex RF-WPT systems and make charging policies more adaptive to application semantics. In this sense, GenAI can be investigated for charging-load forecasting, scenario generation, smart charging, routing, and grid-aware scheduling~\cite{zhang2025ioev,mohammadabadi2024ev}. Although RF-WPT has distinct physical constraints, the same central idea is that when demand is stochastic, context-dependent, and only partially observable, 
GenAI models can improve resource allocation. 

Overall, GenAI can be viewed as an uncertainty-aware modeling layer for charging scheduling. Its main advantage lies in capturing one-to-many mappings between context and future energy needs, generating realistic scenarios for optimization, and enabling proactive, robust, and context-aware schedulers. 
\section{Case Study}

\begin{figure*}[t]
    \centering
    \includegraphics[width=0.95\linewidth]{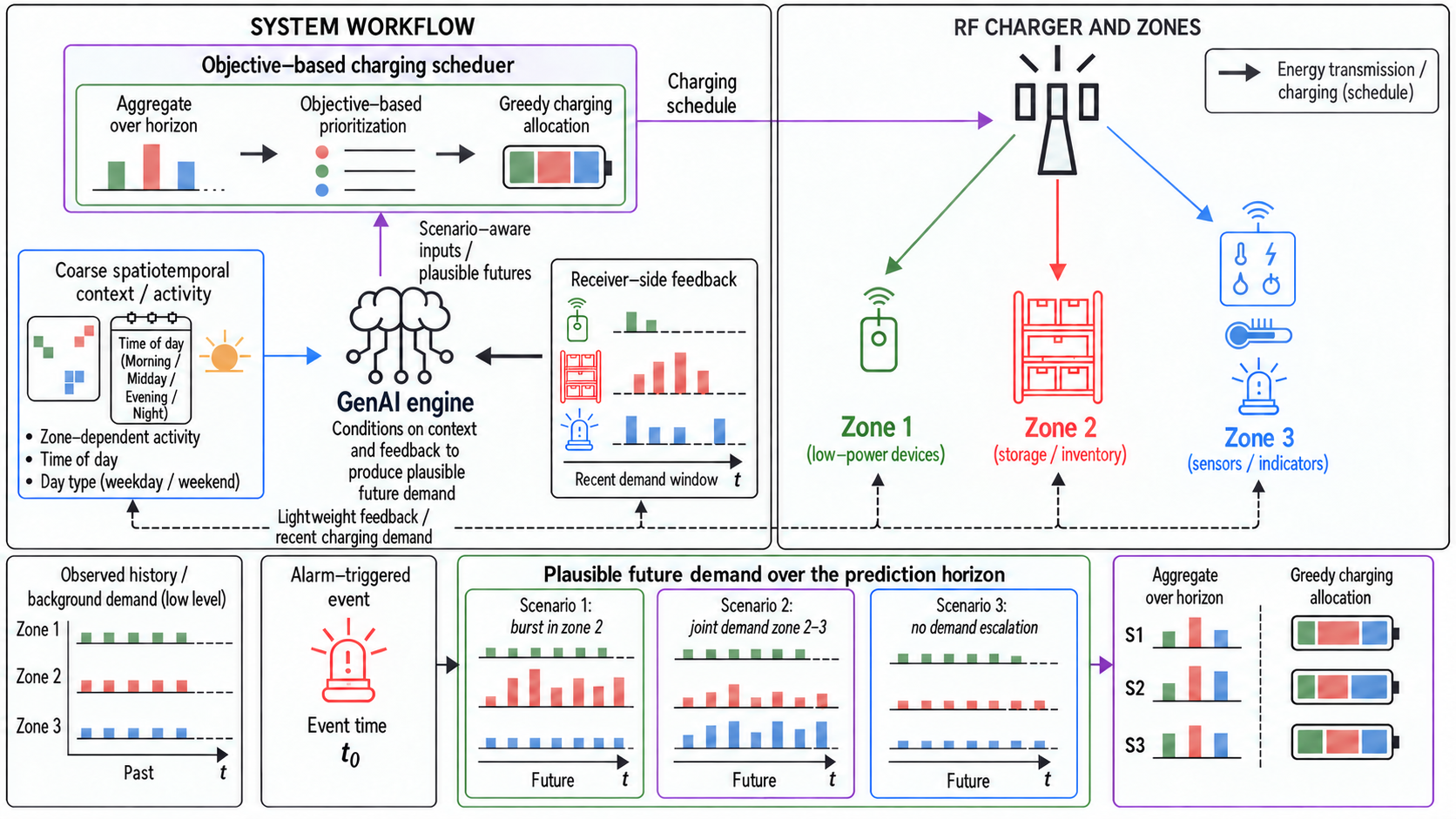}
    \caption{Conceptual illustration of the case study. Coarse spatiotemporal context and lightweight receiver-side feedback are used to condition the GenAI engine. Based on the same observed context, multiple plausible near-future charging-demand scenarios may arise, reflecting the one-to-many nature of alarm-triggered demand surges. The GenAI engine captures this uncertainty and provides scenario-aware inputs to the objective-based charging scheduler, which then determines the RF charger's charging decision across the receiver zones.}
    \label{fig:case_study_example}
\end{figure*}

To make the discussion concrete, we consider a warehouse-style RF-WPT setup, where an RF charger serves multiple receivers in different operational zones. The system usually experiences low background charging demand, while irregular alarm-triggered events create short-term demand surges for an unknown subset of receivers, e.g., due to workload shifts, urgent activity, or localized operational changes. When an alarm occurs, the RF charger must distribute its charging budget over the future window. It obtains a short recent history of receiver charging deficits through lightweight feedback, e.g., uplink signaling or backscatter reporting \cite{AIOTsurvey}, together with contextual descriptors, e.g., warehouse zone, daytime period, and day type. These observations constrain but do not uniquely determine the future demand, so the RF charger operates under structured yet incomplete knowledge of future.

The physical world reflects both predictable and ambiguous demand. Some receivers exhibit regular background loads strongly coupled to context, whereas others form an event-driven branch whose future demand follows a continuous latent process. The resulting charging hotspot may shift across neighboring receivers, broaden or narrow spatially, move within the prediction horizon, and vary in intensity across realizations. Thus, similar observed contexts can still yield different future charging patterns, with a deterministic predictor returning a single future map, while a generative model represents multiple plausible futures. This is operationally important because forecasting errors can direct energy to the wrong receivers while critical devices remain underpowered.

\subsection{Scenario Generation and Scheduling}

{We compare three learned models: i) a conditional diffusion model trained on noisy versions of the future trajectory generating multiple plausible future samples; ii) a convolutional neural network (CNN) that produces a single future charging map; and iii) a deep-ensemble baseline (ENS-CNN), formed by independently initialized and trained CNN predictors whose outputs provide a non-generative distribution of candidate future demands.} The input to learned predictors consists of two components: i) the recent receiver-wise charging history, and ii) coarse contextual indicators describing the current operating zone, the daytime period, and day type (weekday/weekend). {The learned models are trained for 200 epochs on 2,500 alarm events
and evaluated on 20 independently generated 1200-step timelines, with
alarms triggered with probability 0.3 per step. The evaluation is repeated
over 10 independent seeds.}

{At each charging decision, the scheduler aggregates the predicted
charging-demand map over the six-slot horizon to obtain receiver-level
demand totals and allocates a fixed total budget of 30 energy units for
the entire horizon, once across the eight receivers. For Diffusion and
ENS-CNN, 200 future-demand samples are provided per event, whereas the
CNN provides a single deterministic horizon-total demand vector. The
physical simulator directly provides 200 simulator-generated
demand samples to the same scheduler under identical allocation
constraints. A receiver deficit is its unmet horizon-total demand after
allocation. We consider four criteria: average deficit across all receivers; worst deficit, given by the largest deficit among the
five event-driven receivers; top-2 deficit, given by the sum of their two
largest deficits; and the 90th-percentile top-2 deficit across the
sampled scenarios. For sampling-based methods, the first three criteria
are averaged across the 200 scenarios, whereas the last captures
unfavorable but plausible outcomes. The allocation is constructed
greedily in unit-energy increments such that each unit is assigned to
the receiver producing the largest reduction in the selected criterion
until the budget is exhausted. Together with Fig.~\ref{fig:case_study_example}, this specifies the
complete flow from observed inputs and demand scenarios to the
constrained charging allocation and its evaluation.} We also consider two non-learning baselines: a context-based heuristic that allocates part of the budget according to predictable contextual tendencies and distributes the rest according to a context-dependent branch profile; and a uniform allocator that spreads the budget uniformly. These simple baselines represent two common fallback behaviors: i) exploiting only coarse prior knowledge, or ii) avoiding commitment by spreading energy.

Note that this case study intentionally abstracts from detailed physical-layer modeling to isolate the value of GenAI for uncertainty-aware charging scheduling. Thus, the simulator focuses on background demand, alarm-triggered intervention, coarse context, limited charging budget, and multimodal future demand. {The reported gains therefore reflect scheduling-level improvements rather than end-to-end RF-WPT performance, since propagation, beamforming, RF-to-DC conversion, and other physical-layer effects may alter the absolute charging outcomes \cite{AIOTsurvey}.} Fig.~\ref{fig:case_study_example} illustrates the workflow and an example of the one-to-many mapping in plausible future outcomes.

\subsection{Charging Performance}

\begin{figure}[t]
    \centering
    \includegraphics[width=0.95\linewidth]{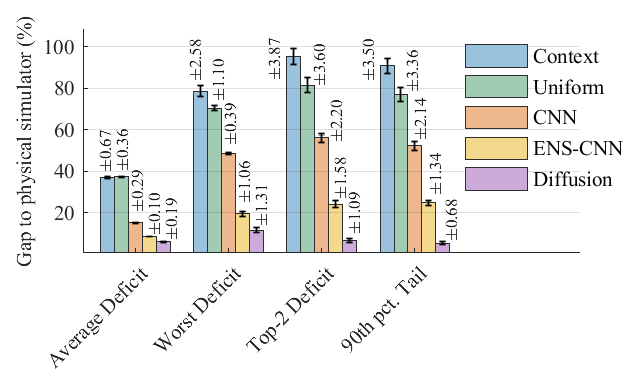}
    \caption{{Average gap to the physical-simulator reference across the four evaluation objectives. Each gap is normalized by the corresponding same-seed physical-simulator reference value. Results are averaged over 10 independent random seeds, with a 95\% confidence interval (CI).}}
    \label{fig:gain_results}
\end{figure}


\begin{figure}[t]
    \centering
    \includegraphics[width=0.95\linewidth]{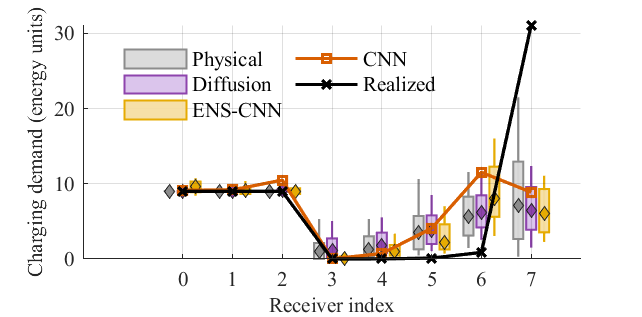}
    \caption{{Representative event-level receiver-wise demand distribution. 
    For the physical simulator, Diffusion, and ENS-CNN, boxes show the 
    25th--75th percentile range, vertical lines the 10th--90th percentile range, and 
    diamonds the median. CNN point predictions and realized demand are also shown.}}
    \label{fig:demand_results}
\end{figure}

\begin{figure}[t]
    \centering
    \includegraphics[width=0.95\linewidth]{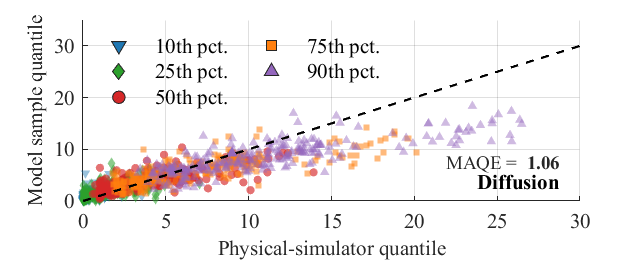}\\
    {(a)}\\[-0.5mm]
    \vspace{0.8mm}
    \includegraphics[width=0.95\linewidth]{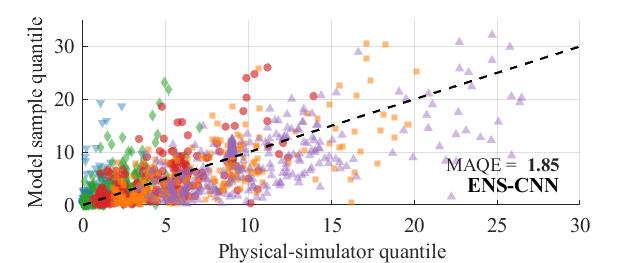}
    {(b)}\\[-0.5mm]
    \caption{{Predicted-vs-physical quantile match for the two sampling-based methods. Model-sampled quantiles (10th/25th/50th/75th/90th percentile of receiver charging demand) are plotted against the corresponding physical-simulator quantiles, pooled across evaluated events and receivers, for (a) Diffusion and (b) ENS-CNN. The dashed diagonal denotes perfect agreement. }}
    \label{fig:case_study_results_dist}
\end{figure}

{Fig.~\ref{fig:gain_results} reports the average gap of each method to the physical-simulator reference. ENS-CNN substantially reduces the gap relative to the deterministic CNN, confirming that providing the scheduler with multiple plausible predictions is itself beneficial. However, Diffusion remains consistently closest to the physical-simulator reference, particularly for the deficit-based objectives. This means that the observed benefit does not arise solely from replacing a point prediction with multiple candidate forecasts, but also from more
faithfully representing the underlying demand distribution.}

{Fig.~\ref{fig:demand_results} provides an event-level view of this behavior.} The unequal receiver-wise demand levels are a direct consequence of the synthetic world design. The first receivers mainly reflect predictable background activity, whereas the latter receivers belong to the event-driven branch. The important observation is that, for the devices with predictable behavior, all learning approaches can estimate the demand reasonably well. In contrast, the physical conditional demand is spread across several receivers in the event-driven branch. {Diffusion more closely follows this spread, while ENS-CNN generally produces a narrower predictive range and the deterministic CNN commits to a single receiver-wise pattern. The medians shown in Fig.~\ref{fig:demand_results} are point summaries of the sampled distributions, while collapsing a sampled distribution to such a point representation discards the uncertainty information contained in its spread and makes the representation closer in nature to the deterministic CNN.}

{To assess model behavior independently of downstream scheduling performance, Fig.~\ref{fig:case_study_results_dist} directly compares the predicted and physical-simulator demand quantiles. Diffusion follows the ideal quantile-matching diagonal more closely across the considered range, whereas ENS-CNN shows larger deviations, particularly toward the higher-demand quantiles. Quantitatively, Diffusion achieves a mean absolute quantile error (MAQE) of 1.06 energy units, compared with 1.85 for ENS-CNN, which shows that Diffusion more faithfully represents the physical-simulator conditional demand distribution.}

\section{Conclusions and future research directions}

RF-WPT scheduling in multi-receiver low-power IoT networks requires the RF charger to determine how limited energy resources should be delivered under uncertain and time-varying environmental and receiver conditions. In this article, we argued that GenAI is promising for this setting because it can represent multiple plausible charging scenarios rather than collapsing uncertainty into a single forecast. This can provide richer support for receiver prioritization, charging timing, and resource allocation. At the same time, realizing this vision raises several key challenges and future research directions:
\begin{itemize}

    \item \textbf{Scalable use of generated scenarios:} Multiple generated scenarios can expose the scheduler to uncertainty-aware information, but also raise the question of how this information should be exploited efficiently. Promising directions include scenario reduction, clustering-based scenario selection, uncertainty-aware ranking, and distributionally robust scheduling, where the scheduler acts on a compact set of informative scenarios rather than all generated samples. Such approaches can improve charging decisions without excessive computational complexity.

    \item \textbf{Generalization to unseen scenarios:} A major challenge is ensuring that generative models remain reliable when the contextual information distribution shifts from the training data. Future work should investigate robust training strategies, online adaptation, and domain transfer methods to keep GenAI-assisted schedulers effective under evolving operating conditions.

    \item \textbf{Data efficiency and trustworthy synthetic data:} Training effective GenAI models requires rich representative data. In practice, such datasets may be limited, while synthetic data must be carefully validated to avoid unrealistic patterns or bias. Promising directions include data-efficient fine-tuning with real-world data, physics-informed generation, and validation mechanisms comparing synthetic scenarios against observed physical statistics. These are needed to make synthetic scenario generation useful and trustworthy for RF-WPT scheduling.

    \item \textbf{Lightweight deployment and real-time inference:} GenAI-assisted scheduling may introduce additional training, inference, and implementation costs. These overheads must remain justified by the resulting performance gains, especially in energy-constrained IoT systems. Promising directions include model compression, knowledge distillation, event-triggered inference, and hierarchical frameworks that help the GenAI engine produce compact outputs and meet real-time scheduling requirements with manageable overhead.

    \item \textbf{Cross-layer integration with the physical layer:} Scheduler-level decisions must ultimately be translated into physical-layer actions. Future research should study how scenario-aware GenAI models can be integrated with RF-WPT physical-layer control without losing tractability or robustness, e.g., through beam selection, waveform design, or power allocation.

    \item \textbf{Risk-aware and reliable charging decisions:} GenAI outputs are integrated into the RF-WPT optimization loop, affecting device availability and service continuity. Thus, charging decisions must remain reliable on average and under rare yet important events. This calls for further research on uncertainty calibration, risk-aware scheduling, and reliable decision support mechanisms that prevent undercharging of critical devices.

\end{itemize}

{Overall, GenAI offers a promising direction for proactive
and context-aware RF-WPT scheduling, but its practical value depends on whether future designs can remain scalable,
data-efficient, robust, and lightweight enough for deployment.}


\bibliographystyle{IEEEtran}
\bibliography{IEEEabrv,references}

\section*{Biographies}

\noindent\textbf{Amirhossein Azarbahram} [S] is a postdoctoral researcher at the Centre for Wireless Communications (CWC), Oulu, focused on integrated sensing and communications and AI/GenAI-native wireless systems. 

\noindent\textbf{Osmel M. Rosabal} [M'25, S'21] is a postdoctoral researcher, focused on RF localization and sensing in IoT, at the CWC, University of Oulu, Finland.

\noindent\textbf{David E. Ruiz-Guirola} [S'19, M'25] is a postdoctoral researcher focused on sustainable IoT, machine learning, and traffic prediction at the University of Oulu, Finland.

\noindent\textbf{Melike~Erol-Kantarci} is Canada Research Chair in AI-enabled Next-Generation Wireless Networks and Full Professor at the School of Electrical Engineering and Computer Science at the University of Ottawa. She is also Strategic Product Manager for AI in RAN at Ericsson. She was an IEEE ComSoc Distinguished Lecturer between 2020-2023. Dr. Erol-Kantarci is an IEEE Fellow.

\noindent\textbf{Kaibin Huang} is the Philip K. H. Wong Wilson K. L. Wong Professor in Electrical Engineering and the Department Head at the Department of Electrical and Computer Engineering, The University of Hong Kong (HKU), Hong Kong. He is a member of the Engineering Panel of the Hong Kong Research Grants Council. He was an IEEE Distinguished Lecturer (2020--2022). He is Croucher Senior Research Fellow (2026), Fellow of the IEEE (2021), and Fellow of the U.S. National Academy of Inventors (2024).

\noindent\textbf{Onel L\'opez} [S'17, M'20, SM'24] is an Associate Professor of wireless communications engineering, focused on sustainable IoT connectivity, at 6G Flagship, University of Oulu, Finland.



\end{document}